# Toward a Market Model for Bayesian Inference

David M. Pennock and Michael P. Wellman
University of Michigan
Artificial Intelligence Laboratory
1101 Beal Avenue
Ann Arbor, MI 48109-2110 USA
{dpennock,wellman}@umich.edu

## Abstract

We present a methodology for representing probabilistic relationships in a general-equilibrium economic model. Specifically, we define a precise mapping from a Bayesian network with binary nodes to a market price system where consumers and producers trade in uncertain propositions. We demonstrate the correspondence between the equilibrium prices of goods in this economy and the probabilities represented by the Bayesian network. A computational market model such as this may provide a useful framework for investigations of belief aggregation, distributed probabilistic inference, resource allocation under uncertainty, and other problems of decentralized uncertainty.

## 1 GENERAL MOTIVATIONS

A principled market model for Bayesian inference would potentially address a variety of important and interesting problems of distributed uncertain reasoning. Although the particular contributions of our work to date do not deliver on these general problems, we present the big picture at the outset as underlying motivation for our specific developments.

There are several reasons one might want to build markets for probabilistic reasoning. Researchers in uncertain reasoning are likely to be acquainted with the first two aims enumerated in sections below. The third may be more familiar to those with some background in economics. The approach we follow in this work owes much to both economic theory (especially general equilibrium under uncertainty) and uncertain reasoning technology (especially Bayesian networks).

### 1.1 AGGREGATING BELIEFS

Given several agents with incompatible beliefs, how can we aggregate their individual beliefs into a characterization of the group's beliefs? This is a classical question in uncertain reasoning, one that has eluded definitive answers despite the research attention it has attracted (Genest and Zidek 1986). Although we doubt that definitive solutions are forthcoming from any quarter, we point out that market mechanisms of various sorts are widely used in uncertain contexts, and their function as aggregators of belief are well-recognized. For example, the price of a stock represents the "market evaluation" of the expected present value of future dividends, and odds in a horse race aggregate the bettors' beliefs about the winning horse's identity.

Despite their commonality and well-developed underlying theory, there appear to have been few attempts in the uncertain reasoning community at principled application of market ideas for belief aggregation. Specifically, to our knowledge, nobody has proposed a comprehensive market architecture for belief aggregation.[1]

### 1.2 DISTRIBUTED PROBABILISTIC INFERENCE AND DECISION UNDER UNCERTAINTY

Belief aggregation is an instance of the more general problem of coordinating belief and decision among a collection of agents. If probabilistic information is distributed across multiple sources, we face the problem of combining this information to address queries and decisions dependent on disparate pieces.

---

[1] though Hanson (1991) has presented an informed argument for, and preliminary investigation of, the idea. In particular, he has advocated setting up a market in scientific claims, where prices summarize consensus opinions about important research questions (Hanson 1995). A prototype of this "Idea Futures" market is in operation on the World-Wide Web, at http://if.arc.ab.ca/IF.shtml.

Much work in uncertain reasoning is amenable to distribution, and indeed, *belief propagation* mechanisms are often specifically designed for this purpose. Market mechanisms likewise are directly geared toward decentralization, in particular for situations where the participating modules are viewed as rational agents (Wellman 1995). Understanding the decentralization in markets for uncertain propositions can perhaps lead to new distributed reasoning schemes, or improvements in existing schemes.

### 1.3 RESOURCE ALLOCATION UNDER UNCERTAINTY

Real markets involve substantial uncertainty. Agents making decisions in markets often have to commit to consumption or production plans without an ability to predict perfectly the outcomes of those plans. This state of uncertainty may be shared by the agents, or each may have different uncertain beliefs. The resource allocation problem under uncertainty includes allocation of risks, distributing the consequences of this uncertainty throughout the system.

Economics has developed a substantial body of theory addressing uncertainty in markets. The standard theory of general equilibrium under uncertainty posits *contingent goods*—goods that are available only if particular uncertain events are realized. A *security* is a particular kind of contingent good with varying monetary (or real-good) payoffs based on the revealed value of some uncertain state. One of the key results of this theory is that an economy with a sufficient set of securities markets possesses the classical properties of competitive equilibrium (i.e., existence, uniqueness, Pareto optimality) under conditions identical to that for the case of certainty (essentially, convexity and continuity of preferences and technologies). See a standard text on microeconomic theory (e.g., Mas-Colell, et al. (1995)) for the precise statement and detailed development of these concepts.

To be *sufficient*, a set of securities markets must span the set of possible states of nature, $\Omega$. This requires in general $O(|\Omega|)$ securities. If the set of markets is *incomplete*, allocations dictated by the price system may be inefficient (i.e., Pareto dominated), even if the economy is otherwise well-behaved.

Of course, research in uncertain reasoning is largely concerned with representing beliefs over $\Omega$, and one of the major ideas is that structure in this set may lead to more compact and otherwise advantageous specifications. In particular, we typically structure $\Omega$ by factoring it into a product of random variables, and achieve savings in encoding (e.g., in Bayesian networks) by exploiting independence among the random variables.

A natural question, then, is whether ideas about structuring probabilistic relations among random variables can be exploited in the design of configurations of securities markets. We believe the work presented below provides evidence for an answer in the affirmative.

## 2  OVERVIEW

In this paper we present one approach toward building a market system for Bayesian inference. We propose a market structure to represent a joint probability distribution over a set of binary random variables. Specifically, we delineate a precise mapping from a Bayesian network to an economy, where consumers and producers exchange goods representing uncertain propositions. The resulting system, called *MarketBayes*, is shown to effectively "compute" the same probabilistic information as the original Bayesian network. While this paper focuses on the particulars of the MarketBayes construction, our aim is not the constructed market system *per se*, but rather its use as a foundation for future investigations of the broader issues entertained above.

In the next section, we provide some general background on the microeconomic framework employed. In Section 4, we introduce the MarketBayes model, beginning with the basic agents and goods that comprise our market structure. The complete mapping from Bayesian networks to market structures is presented in Section 5, along with some theorems characterizing the correspondence between the resulting market price system and the original joint probability distribution. We illustrate the method further by example in Section 6, and conclude in the subsequent section with a discussion and assessment of our results.

## 3  ECONOMIC FUNDAMENTALS

### 3.1  GENERAL EQUILIBRIUM FRAMEWORK

A *market price system* is defined by a set of $K$ *goods* indexed $1, \ldots, K$ and a *price vector* $\mathbf{p} = (p_1, \ldots, p_K)$ that associates a price with each good. The system requires that all goods be exchanged in proportion to their relative prices.

Goods are exchanged by two types of *agents*—consumers and producers. *Consumer* agents receive value from direct consumption of the goods (metaphorically, they "eat" the goods). Let $\mathbf{x} = (x_1, \ldots, x_K)$ denote a *consumption bundle* where each $x_i \in \Re_+$ specifies the quantity of good $i$ consumed. The consumption bundles are ranked according to preference by the consumer's *utility function* $u(\mathbf{x})$:

$\Re^K_+ \to \Re$. Consumers also start with an initial allocation of the goods, termed their *endowment* and denoted by $e = (e_1, \ldots, e_K)$. The consumer's objective is to choose an affordable bundle of goods, $\mathbf{x}$, so as to maximize its utility. A bundle is affordable if its total cost at the going prices does not exceed the value of the consumer's endowment at the same prices. The consumer's choice can thus be expressed as the following constrained optimization problem:

$$\max_{\mathbf{x}} u(\mathbf{x}) \text{ s.t. } \mathbf{p} \cdot \mathbf{x} \leq \mathbf{p} \cdot \mathbf{e}. \tag{1}$$

Agents of the second type, *producers*, extract value from goods by transforming them into other goods, and selling their product in the market. A producer's ability to transform goods is defined by its *technology*, $Y \subset \Re^K$, which specifies the set of feasible production vectors. If $\mathbf{y} = (y_1, \ldots, y_K)$ is feasible, then the producer is capable of transforming bundles of *input* goods (goods $i$ for which $y_i < 0$) into bundles of *output* goods ($y_i > 0$), in respective amounts $|y_i|$.

Unlike the consumer, a producer has no preferences in the sense of agent-specific desires. Rather, we assume that a producer selects its production activity solely according to *profit*—the difference between the value of its output and the cost of its input, evaluated at a given set of prices. The producer's constrained optimization problem can be expressed succinctly as

$$\max_{\mathbf{y}} \mathbf{p} \cdot \mathbf{y} \text{ s.t. } \mathbf{y} \in Y. \tag{2}$$

An agent is *competitive* if it takes prices as given, ignoring the potential effect of its own choices on resulting prices. The agent definitions above assume competitive behavior, in that the prices are treated as parameters of the respective optimization problems. Note that only relative prices matter; behavior is unchanged if all prices are multiplied by a positive constant. We typically scale prices by designating one good (the first, without loss of generality) as *numeraire*, with a fixed price, $p_1 = 1$.

Consider an economy with consumers indexed $1, \ldots, m$ and producers indexed $m + 1, \ldots, n$. A *competitive equilibrium* for this economy is a set of prices, $\mathbf{p}$, such that all of the goods are in material balance,

$$\sum_{i=1}^{m} \mathbf{x}^i \leq \sum_{i=m+1}^{n} \mathbf{y}^i + \sum_{i=1}^{m} \mathbf{e}^i,$$

where $\mathbf{x}^i$ and $\mathbf{y}^i$ denote the solutions of consumer or producer $i$'s respective optimization problem, as defined above, at prices $\mathbf{p}$.

## 3.2 CONTINGENT GOODS

In economic theory, uncertainty is addressed by introducing contingent goods, whose tangible realization depends on some uncertain event. In the standard model, trading is divided into two periods. In the first period (before uncertainty is resolved) agents trade exclusively in securities. After uncertainty is resolved, agents trade in the real goods, using income from their securities holdings.

Since we are developing a market for uncertainty in the abstract (i.e., no explicit real goods), we consider a one-stage model, with the securities treated as if normal goods. Consider a security that pays one "dollar" if an uncertain proposition $a$ is true, and nothing otherwise. If an agent is risk neutral for dollars, then its valuation of this good is exactly that of having $\Pr(a)$ dollars. If *dollar* is the numeraire, then the price at which the agent would be indifferent between buying one unit of the contingent good and one dollar is exactly $\Pr(a)$. Thus, the equilibrium price in this instance corresponds with the probability.

More generally, we interpret the price of a good as "the market probability" for that good. If participating consumers have different assessments of the probability, then their relative wealth and risk aversion will determine their influence on the equilibrium price.

The market structure we define below is comprised entirely of contingent goods of the form: "$1 if $a$", for propositions of interest $a$. We specify preferences over these goods directly, rather than assigning beliefs in the propositions and preferences for dollars. Our selection of which propositions to include and our definition of the participating agents are designed to achieve an equilibrium where the prices correspond to a specified probability distribution.

## 4 MARKETBAYES BUILDING BLOCKS

The *MarketBayes* model is a particular approach we have developed to represent joint probability distributions over sets of binary random variables in terms of general-equilibrium economic systems. The market representation is designed to exploit conditional independencies among propositions, based on the structure of a Bayesian network. Specifically, we present a mapping from a Bayesian network with binary propositional nodes to an "equivalent" configuration of goods, consumers, and producers. In this section, we present the basic economic constructs that form the building blocks of the MarketBayes economy. Section 5 presents the complete mapping, and characterizes the correspondence between the resulting market price system and the original Bayesian network.

We represent propositions formed from conjunctions of nodes in the network as contingent goods in the economy. Let $a$ be a proposition, and $\langle a \rangle$ the corresponding good in the market model. We also use the notation $\langle a \rangle$ as a variable denoting the price of the good; the specific interpretation is resolvable in context. By design the equilibrium price of the good should equal the probability of the corresponding proposition, i.e., $\langle a \rangle = \Pr(a)$. We define next the MarketBayes consumers and producers employed in pursuit of this equilibrium behavior.

## 4.1 THE CONSUMERS

In a Bayesian network, the basic unit of information is a conditional probability. For example, a node $a_2$ with sole predecessor $a_1$ is accompanied by the information $\Pr(a_2|a_1) = k$ where $k$ is some probability. Using the definition of conditional probability, the same equation can be rewritten as $\Pr(a_1 a_2) = k \Pr(a_1)$. In the MarketBayes economy, we wish to enforce the same ratio between the prices of the goods:

$$\langle a_1 a_2 \rangle = k \langle a_1 \rangle. \quad (3)$$

Our approach to maintaining this relation is to introduce a consumer that considers the relative value of the good $\langle a_1 a_2 \rangle$ to be $k$ times the value of the good $\langle a_1 \rangle$. If the ratio of the prices diverges from $k$, the consumer will buy or sell accordingly, tending to drive the ratio toward $k$.

The MarketBayes model employs CES (constant elasticity of substitution) consumers for this purpose. The CES utility function for two goods takes the form[2]

$$u(x_1, x_2) = \left( \alpha_1 (x_1)^{\frac{\sigma-1}{\sigma}} + \alpha_2 (x_2)^{\frac{\sigma-1}{\sigma}} \right)^{\frac{\sigma}{\sigma-1}}, \quad (4)$$

where the $\alpha_i$ are coefficients dictating the relative values of the two goods, and $\sigma$ is a global substitution parameter dictating the degree to which consumption in one good (at proportions dictated by the $\alpha_i$) can substitute for the other.

Let $p_1$ and $p_2$ be the prices of the two goods. The consumer's optimization problem (1), as usual, is to maximize its utility function (4) subject to its budget constraint. For CES consumers this problem has a closed-form solution:

$$x_1(p_1, p_2) = \frac{\alpha_1^\sigma (p_1 e_1 + p_2 e_2)}{p_1^\sigma (\alpha_1^\sigma p_1^{1-\sigma} + \alpha_2^\sigma p_2^{1-\sigma})} \quad (5)$$

$$x_2(p_1, p_2) = \frac{\alpha_2^\sigma (p_1 e_1 + p_2 e_2)}{p_2^\sigma (\alpha_1^\sigma p_1^{1-\sigma} + \alpha_2^\sigma p_2^{1-\sigma})} \quad (6)$$

---
[2] CES forms are commonly employed in general equilibrium modeling (Shoven and Whalley 1992), due to their flexibility and convenient analytical properties.

To implement an equation of the form (3), we introduce a CES consumer interested in the two goods $\langle a_1 a_2 \rangle$ and $\langle a_1 \rangle$. The consumer is endowed with an equal amount $e_1 = e_2 = e$ of each good (the exact value does not matter for the current purpose, as long as $e > 0$). By setting $\alpha_1 = k$ and $\alpha_2 = 1$, we encode the desired conditional probability. Although the relation is strictly enforced (according to Theorem 1 below) only as $\sigma \to \infty$, we have found in practice that convergence to the correct price ratio typically obtains for values of $\sigma > 4$ or so.

**Theorem 1** *Let $\langle a_1 a_2 \rangle^*$ and $\langle a_1 \rangle^*$ be equilibrium prices for the two goods in an economy containing the CES consumer defined above, with $\sigma \to \infty$. If $\lim_{\sigma \to \infty} \frac{\langle a_1 a_2 \rangle^*}{\langle a_1 \rangle^*}$ is finite and bounded away from zero, then it is $k$.*

**Proof** (sketch). In competitive equilibrium, by definition, the consumer is solving its optimization problem. The first-order conditions for that optimization problem dictate that the marginal utility per unit price be constant across goods. In particular,

$$\frac{\frac{\partial u}{\partial x_1}}{\langle a_1 a_2 \rangle^*} = \frac{\frac{\partial u}{\partial x_2}}{\langle a_1 \rangle^*}.$$

Substituting the marginal utilities and rearranging,

$$\begin{aligned}\frac{\langle a_1 a_2 \rangle^*}{\langle a_1 \rangle^*} &= \frac{\left( k(x_1)^{\frac{\sigma-1}{\sigma}} + (x_2)^{\frac{\sigma-1}{\sigma}} \right)^{\frac{1}{\sigma-1}} k(x_1)^{-\frac{1}{\sigma}}}{\left( k(x_1)^{\frac{\sigma-1}{\sigma}} + (x_2)^{\frac{\sigma-1}{\sigma}} \right)^{\frac{1}{\sigma-1}} (x_2)^{-\frac{1}{\sigma}}} \\ &= \frac{k(x_1)^{-\frac{1}{\sigma}}}{(x_2)^{-\frac{1}{\sigma}}},\end{aligned}$$

which approaches $k$ as $\sigma \to \infty$, as long as the $x_i$ are bounded away from zero. For CES consumers, this will be true as long as the price ratio is finite and bounded away from zero. Thus, $\langle a_1 a_2 \rangle^* = k \langle a_1 \rangle^*$, which is Equation (3) exactly. □

The alert reader may observe that the same result could have been obtained more directly using the linear utility function, $u(x_1, x_2) = k x_1 + x_2$. Indeed, the CES utility function approaches linearity in the limit. However, with linear utility the equilibrium would be more fragile—two consumers with different $k$ would prevent existence. Moreover, for linear utility functions the optimal demand function is discontinuous in prices, and reaching equilibrium through a distributed, incremental bidding process becomes more difficult.

Note that Theorem 1 requires only that there exist one such CES consumer. It is true vacuously when there is no equilibrium. For situations with more than one CES consumer connecting the same pair of goods with differing $k$ values, the result still holds because we take the $\sigma$ for only one of them to infinity.

## 4.2 THE PRODUCERS

Whereas the role of consumers in a MarketBayes economy is to encode conditional probabilities, we employ producers to implement identities of probability theory. Producers act as *arbitrageurs*, converting between logically equivalent bundles of goods.[3] For example, the propositional identity

$$a_1 \equiv a_1 a_2 \vee a_1 \bar{a}_2$$

can be expressed in a producer with the technology to convert a unit of $\langle a_1 \rangle$ into one unit each of $\langle a_1 a_2 \rangle$ and $\langle a_1 \bar{a}_2 \rangle$, or *vice versa*. This producer's technology exhibits *constant returns to scale*, that is, it can perform this transform at any volume level.

The corresponding probabilistic identity, $\Pr(a_1) = \Pr(a_1 a_2) + \Pr(a_1 \bar{a}_2)$, can likewise be expressed in terms of a constraint on prices of goods:

$$\langle a_1 \rangle = \langle a_1 a_2 \rangle + \langle a_1 \bar{a}_2 \rangle. \tag{7}$$

The arbitrageur effectively enforces this equation by its bidding policy. If the price $\langle a_1 \rangle$ diverges from the sum $\langle a_1 a_2 \rangle + \langle a_1 \bar{a}_2 \rangle$, the producer can make profits by transforming one side to the other. Its resulting demand behavior will tend to drive the respective input and output prices towards equality.

The producer's goal is to maximize profits (2). For the producer associated with the identity above, the profits when transforming $y$ units of $\langle a_1 \rangle$ into $y$ each of $\langle a_1 a_2 \rangle$ and $\langle a_1 \bar{a}_2 \rangle$ (note that if $y$ is negative, the transformation goes the other way) are simply

$$y\left(\langle a_1 a_2 \rangle + \langle a_1 \bar{a}_2 \rangle - \langle a_1 \rangle\right). \tag{8}$$

**Theorem 2** *Let $\langle a_1 \rangle^*$, $\langle a_1 a_2 \rangle^*$, and $\langle a_1 \bar{a}_2 \rangle^*$ be the equilibrium prices for the three goods in an economy containing the arbitrage producer defined above. Then $\langle a_1 \rangle^* = \langle a_1 a_2 \rangle^* + \langle a_1 \bar{a}_2 \rangle^*$.*

**Proof.** Competitive producers must be maximizing profits in equilibrium. But the profit function (8) has a bounded maximizer $y$ (finite production) only if Equation 7 holds. □

Note that the producer always makes zero profit in equilibrium (this is true in general for competitive, constant-returns producers). Results analogous to Theorem 2 can be derived for arbitrageurs representing identities of the form (7) but with arbitrary numbers of propositions on the right-hand side.

---

[3]Our expression of the laws of probability in arbitrageurs can be viewed as a direct embodiment of Nau and McCardle's argument (1991) that all rationality or coherence principles ultimately reduce to "no-arbitrage" postulates.

## 5 THE MARKETBAYES SYSTEM

In this section we piece together the individual components described above into an interconnected market price system. We describe a general mapping from a Bayesian network with binary nodes to a MarketBayes configuration of goods, consumers, and producers. We show how the economy effectively represents the same information as the Bayesian network.

We are interested in three general properties that such a mapping may possess.

**Property 1 (Existence)** *There exists a competitive equilibrium in the MarketBayes economy such that the price of each good equals the probability of the corresponding proposition in the Bayesian network.*

**Property 2 (Uniqueness)** *There is a unique competitive equilibrium in the MarketBayes economy, satisfying the conditions of Property 1.*

**Property 3 (Convergence)** *The unique competitive equilibrium of the MarketBayes economy can be derived via an iterative, distributed bidding process, such as tatonnement or variants (Mas-Colell, Whinston, and Green 1995; Cheng and Wellman 1996).*

Each property subsumes the previous and, in general, the properties are successively harder to verify. In the following sections, we construct the MarketBayes economy incrementally in two stages. After stage one, we prove that Property 1 holds; after stage two we prove that Property 2 holds. The goods in the economy—specified in Section 5.1—remain constant across both stages. In Section 5.2, we define stage one of the mapping, a set of consumers sufficient to establish the Existence Property for arbitrary Bayesian networks. In Section 5.3 we define stage two of the mapping, adding producers to the economy to ensure the Uniqueness Property for a non-restrictive class of Bayesian networks. We do not yet have a general proof of the Convergence Property (except for complete graphs, not presented), but we conjecture that our price adjustment algorithm (Cheng and Wellman 1996) does converge for a broad class of MarketBayes economies. Our computational experience thus far supports this conjecture, and in Section 6 we present a concrete example that does indeed converge when implemented as a computational economy.

### 5.1 THE GOODS

The goods in a MarketBayes economy are conjunctions of literals, each the value of a node in the Bayesian network. However, rather than include all such conjunctions explicitly as goods, we attempt to exploit the independencies present in a (possibly sparse) Bayesian network graph.

Let $a_1, \ldots, a_n$ be nodes in a Bayesian network. Denote the parents of node $a_i$ by $a_{i_1}, \ldots, a_{i_q}$. For each node $a_i$ in the network we add to the economy goods for all $2^{q+1}$ possible conjunctions of $a_i$ and its parents:[4]

$$\langle a_i a_{i_1} \cdots a_{i_q}\rangle, \langle a_i a_{i_1} \cdots \overline{a}_{i_q}\rangle, \ldots, \langle \overline{a}_i \overline{a}_{i_1} \cdots \overline{a}_{i_q}\rangle.$$

We also add to the economy all $2^q$ possible conjunctions of the parents of $a_i$ alone, if these goods are not already included in those defined previously.

$$\langle a_{i_1} \cdots a_{i_q}\rangle, \langle a_{i_1} \cdots \overline{a}_{i_q}\rangle, \ldots, \langle \overline{a}_{i_1} \cdots \overline{a}_{i_q}\rangle$$

Finally we add a single good $\langle T \rangle$ which corresponds to the proposition **true**. $\langle T \rangle$ is the numeraire of the economy and thus its price is 1 by definition.

The economy then consists of $O(2^q n)$ goods, where $q$ is the maximum number of parents for a node and $n$ is the number of nodes.

## 5.2 STAGE ONE: THE CONSUMERS

For each node $a_i$, the given Bayesian network provides us with $2^q$ conditional probabilities, of the form $\Pr(a_i | a_{i_1} a_{i_2} \cdots a_{i_q}) = k$. These conditional probabilities dictate that the following $2^q$ ratios between prices of goods must hold:

$$\begin{aligned}
\langle a_i a_{i_1} a_{i_2} \cdots a_{i_q}\rangle &= k_1 \langle a_{i_1} a_{i_2} \cdots a_{i_q}\rangle \\
\langle a_i a_{i_1} a_{i_2} \cdots \overline{a}_{i_q}\rangle &= k_2 \langle a_{i_1} a_{i_2} \cdots \overline{a}_{i_q}\rangle \\
&\vdots \\
\langle a_i \overline{a}_{i_1} \overline{a}_{i_2} \cdots \overline{a}_{i_q}\rangle &= k_{2^q} \langle \overline{a}_{i_1} \overline{a}_{i_2} \cdots \overline{a}_{i_q}\rangle
\end{aligned} \quad (9)$$

From the above equations, it is trivial to derive the complementary equations that contain $\overline{a}_i$:

$$\begin{aligned}
\langle \overline{a}_i a_{i_1} a_{i_2} \cdots a_{i_q}\rangle &= (1-k_1) \langle a_{i_1} a_{i_2} \cdots a_{i_q}\rangle \\
\langle \overline{a}_i a_{i_1} a_{i_2} \cdots \overline{a}_{i_q}\rangle &= (1-k_2) \langle a_{i_1} a_{i_2} \cdots \overline{a}_{i_q}\rangle \\
&\vdots \\
\langle \overline{a}_i \overline{a}_{i_1} \overline{a}_{i_2} \cdots \overline{a}_{i_q}\rangle &= (1-k_{2^q}) \langle \overline{a}_{i_1} \overline{a}_{i_2} \cdots \overline{a}_{i_q}\rangle
\end{aligned} \quad (10)$$

Root nodes in the Bayesian network can be handled in the same way by considering them to be children of the proposition **true**, with "conditional" probability $\Pr(a_i | \text{true}) = k$.

In the MarketBayes model, CES consumers effectively implement equations of the form (9) and (10). Specifically, we add $2 \cdot 2^q$ consumers to the economy for each node $a_i$—one for each of the equations in (9) and (10). The consumers are instantiated as described in Section 4.1.

---

[4]Though each node may have a different number of parents, we conserve subscripts and use $q$ for the number of parents of the *current* node.

**Theorem 3** *Property 1 (Existence) holds for the mapping from any Bayesian network to the set of goods and consumers defined above.*

**Proof.** We need to show that the probabilities of propositions in the Bayesian network form a (possibly nonunique) price equilibrium in the MarketBayes economy. A set of prices constitutes an equilibrium if the resulting demand is in material balance, that is, there is no *excess demand* in the economy at those prices. Probabilities represented by the Bayesian network obey the ratios described in Equations (9) and (10), by construction. Then it suffices to show that any set of prices that satisfies these equations implies zero excess demand in the economy. Consider the CES consumer associated with the first conditional probability in (9). Let $p_1 = \langle a_i a_{i_1} a_{i_2} \cdots a_{i_q}\rangle$ and $p_2 = \langle a_{i_1} a_{i_2} \cdots a_{i_q}\rangle$ be the prices of the two relevant goods. From (9) we have that $p_1 = k_1 p_2$. The demands for each good (from (5) and (6)) are then:

$$\begin{aligned}
x_1(k_1 p_2, p_2) &= \frac{k_1^\sigma (k_1 p_2 e + p_2 e)}{(k_1 p_2)^\sigma (k_1^\sigma (k_1 p_2)^{1-\sigma} + p_2^{1-\sigma})} \\
&= \frac{k_1^\sigma p_2 e (k_1 + 1)}{k_1^\sigma k_1 p_2 + k_1^\sigma p_2} = \frac{e(k_1+1)}{k_1+1} = e
\end{aligned}$$

$$\begin{aligned}
x_2(k_1 p_2, p_2) &= \frac{k_1 p_2 e + p_2 e}{p_2^\sigma (k_1^\sigma (k_1 p_2)^{1-\sigma} + p_2^{1-\sigma})} \\
&= \frac{p_2 e(k_1 + 1)}{p_2 k_1 + p_2} = \frac{p_2 e(k_1+1)}{p_2(k_1+1)} = e
\end{aligned}$$

In other words, at the specified price ratio, the consumer demands exactly the amounts it is endowed with. Since the same argument applies to every consumer in the economy, the total excess demand must be zero. Thus, any set of prices that satisfies (9) and (10) constitutes a competitive equilibrium for the economy, and the Existence Property is established. □

Note that this result does not require $\sigma \to \infty$, as did Theorem 1. If consumers other than the ones specified in our construction are present, then the network probabilities may not constitute a price equilibrium.

## 5.3 STAGE TWO: THE PRODUCERS

Although the existence of an equilibrium corresponding to the probability distribution is somewhat encouraging, the MarketBayes construction will be of limited use if other, incorrect, probabilities can also represent an equilibrium. In this section we extend the economy to ensure the Uniqueness Property. To do so, we must in general add some consumers and producers to enforce the remaining constraints.

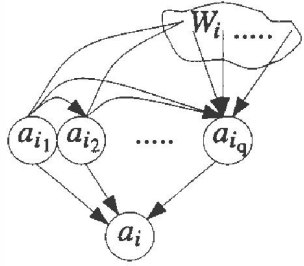

Figure 1: A section of a moral Bayesian network. Undirected links could be oriented either way.

A graph is *moral* if every two parents of a node are connected ("married"), for all nodes in the graph.[5] Forests (collections of trees) are a subset of moral graphs, since every node has at most one parent. Complete graphs are also a subset of moral graphs since *every* two nodes are connected. Any graph can be *moralized* by adding directed links in the graph between every two unconnected parent nodes, making sure to retain the acyclic nature of the network. Adding new directed links introduces new parent relationships which in turn may require additional "moralization" links. Thus, in principle the MarketBayes mapping applies to arbitrary network structures. However, the process of moralization introduces, in the worst case, an exponential number of additional conditional relationships.

As above, let $a_1, \ldots, a_n$ be the nodes of the Bayesian network. Without loss of generality, we can assume the index labels on the nodes are consistent with the partial order represented by the directed graph: if $a_i$ is an ancestor of $a_j$ then $i < j$. Note that $a_1$ is always a root node (no parents) and $a_n$ is always a leaf node (no children). Since the network is moral, the parents of $a_i$—$a_{i_1}, \ldots, a_{i_q}$—form a complete subset of the graph. Furthermore, (taking, again without loss of generality, $i_1 < \cdots < i_q$), the nodes $a_{i_1}, a_{i_2}, \ldots, a_{i_{q-1}}$ must all be parents of the node $a_{i_q}$. Let $W_i$ be the set of parents of the node $a_{i_q}$ that are *not* also parents of $a_i$. This structure is depicted in Figure 1.

The set $W_i$ represents the additional nodes required to specify a joint distribution over the parents of $a_i$. For each node $a_i$, we generate arbitrageurs (according to the scheme of Section 4.2) representing the following probabilistic identities:

$$\begin{aligned}
\langle a_{i_1} a_{i_2} \cdots a_{i_q} \rangle &= \sum_{W_i} \langle W_i a_{i_1} a_{i_2} \cdots a_{i_q} \rangle \\
\langle a_{i_1} a_{i_2} \cdots \overline{a}_{i_q} \rangle &= \sum_{W_i} \langle W_i a_{i_1} a_{i_2} \cdots \overline{a}_{i_q} \rangle \\
&\vdots \\
\langle \overline{a}_{i_1} \overline{a}_{i_2} \cdots \overline{a}_{i_q} \rangle &= \sum_{W_i} \langle W_i \overline{a}_{i_1} \overline{a}_{i_2} \cdots \overline{a}_{i_q} \rangle
\end{aligned} \quad (11)$$

---
[5]The term "moral" in the literature typically refers to the undirected moral relative of a directed graph (Neapolitan 1990). In this paper we use the term as defined above to describe directed graphs only.

The summations are over all possible combinations of the propositions in the set $W_i$. Note that if the set $W_i$ is empty for some $i$, then we need not add any producers to the economy for node $a_i$. If the Bayesian network is a complete graph, then the set $W_i$ will be empty for all $i$; in this case producers are simply not necessary.

**Theorem 4** *Property 2 (Uniqueness) holds for the mapping from any moral Bayesian network to the set of goods, consumers, and producers defined above.*

**Proof.** From Theorems 1 and 2 it is clear that the consumer equations in (9) and (10), along with the producer equations in (11), must hold simultaneously in equilibrium. The probabilities of the propositions in the Bayesian network must satisfy these equations since they correspond directly to the given conditional probabilities plus identities in probability theory. Therefore, we need only show that there is a *unique* set of prices that satisfies this set of equations instantiated for the configuration of this economy. The proof is by induction. Let $G_i$ be the set of goods added to the economy for proposition $a_i$, namely, all possible conjunctions of $a_i$ with its parents plus all possible conjunctions of its parents alone. Mathematically,

$$\begin{aligned}
G_i &= \{\langle a_i a_{i_1} \cdots a_{i_q} \rangle, \ldots, \langle \overline{a}_i \overline{a}_{i_1} \cdots \overline{a}_{i_q} \rangle\} \cup \\
&\quad \{\langle a_{i_1} \cdots a_{i_q} \rangle, \ldots, \langle \overline{a}_{i_1} \cdots \overline{a}_{i_q} \rangle\}.
\end{aligned}$$

- **Base Case.** Define the set $G_0$ to be $\{\langle T \rangle\}$. The good $\langle T \rangle$ is the numeraire and its price is maintained at unity by definition. Thus the price of the good in the set $G_0$ is uniquely determined.

- **Induction.** Assume that all of the prices of goods in the sets $G_0, G_1, \ldots, G_{i-1}$ are uniquely determined. We want to prove that all of the prices of goods in the set $G_i$ are uniquely determined. Consider the general situation as depicted in Figure 1. The prices of the goods in the set $G_{i_q}$ are uniquely determined since $i_q < i$. Among the goods in the set $G_{i_q}$ are the goods in the sets

$$\bigcup_{W_i} \langle W_i a_{i_1} a_{i_2} \cdots a_{i_q} \rangle, \ldots, \bigcup_{W_i} \langle W_i \overline{a}_{i_1} \overline{a}_{i_2} \cdots \overline{a}_{i_q} \rangle,$$

where $\cup_{W_i}$ is the union over all possible conjunctions of the propositions in the set $W_i$. These are exactly the goods on the right hand side of (11). Thus the prices of the goods on the left hand side of (11) must be uniquely determined. These goods are in turn exactly those on the right hand sides of (9) and (10). Thus the prices of the goods on the left hand sides of (9) and (10) must be uniquely determined. The goods on the left hand sides of (9), (10), and (11) are exactly those goods in the set $G_i$. □

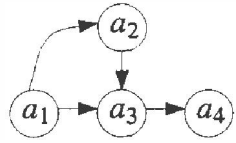

Figure 2: An example Bayesian network.

Note that the price system in an equilibrium Market-Bayes economy is sufficient to recover the complete joint distribution, as it specifies all conditional probabilities in the original Bayesian network.

## 6   AN EXAMPLE ECONOMY

In this section we construct a concrete MarketBayes economy using the technique described in the previous section. We provide empirical verification of the system by reporting results of running the example in an actual computational economy. For this example, the prices of goods converge correctly to the probabilities of the corresponding propositions in the Bayesian network.

The example Bayesian network is pictured in Figure 2. It is already moral, so we need not add any additional links. Let the conditional probabilities associated with the example network be as follows:

$$\begin{array}{ll} \Pr(a_1) = 0.4 & \\ \Pr(a_2|a_1) = 0.2 & \Pr(a_2|\bar{a}_1) = 0.3 \\ \Pr(a_3|a_1 a_2) = 0.11 & \Pr(a_3|\bar{a}_1 a_2) = 0.22 \\ \Pr(a_3|a_1 \bar{a}_2) = 0.33 & \Pr(a_3|\bar{a}_1 \bar{a}_2) = 0.44 \\ \Pr(a_4|a_3) = 0.25 & \Pr(a_4|\bar{a}_3) = 0.85 \end{array}$$

The goods in the economy consist of all combinations of each node with its parents,

$$\begin{array}{llll} \langle a_1 \rangle, & \langle \bar{a}_1 \rangle, & & \\ \langle a_1 a_2 \rangle, & \langle a_1 \bar{a}_2 \rangle, & \langle \bar{a}_1 a_2 \rangle, & \langle \bar{a}_1 \bar{a}_2 \rangle, \\ \langle a_1 a_2 a_3 \rangle, & \langle a_1 a_2 \bar{a}_3 \rangle, & \langle a_1 \bar{a}_2 a_3 \rangle, & \langle a_1 \bar{a}_2 \bar{a}_3 \rangle, \\ \langle \bar{a}_1 a_2 a_3 \rangle, & \langle \bar{a}_1 a_2 \bar{a}_3 \rangle, & \langle \bar{a}_1 \bar{a}_2 a_3 \rangle, & \langle \bar{a}_1 \bar{a}_2 \bar{a}_3 \rangle, \\ \langle a_3 a_4 \rangle, & \langle a_3 \bar{a}_4 \rangle, & \langle \bar{a}_3 a_4 \rangle, & \langle \bar{a}_3 \bar{a}_4 \rangle \end{array}$$

and all combinations of each node's parents alone, if not already included in the group above. In this example, the combinations of the parent of node $a_4$ still need to be added.

$$\langle a_3 \rangle, \langle \bar{a}_3 \rangle$$

Finally we add the numeraire good $\langle T \rangle$ to the economy.

For each conditional probability in the Bayesian network we add a consumer. For this example we have consumers enforcing the following relationships:

$$\begin{array}{ll} \langle a_1 \rangle = 0.4 \langle T \rangle & \\ \langle a_1 a_2 \rangle = 0.2 \langle a_1 \rangle & \langle \bar{a}_1 a_2 \rangle = 0.3 \langle \bar{a}_1 \rangle \\ \langle a_1 a_2 a_3 \rangle = 0.11 \langle a_1 a_2 \rangle & \langle \bar{a}_1 a_2 a_3 \rangle = 0.22 \langle \bar{a}_1 a_2 \rangle \\ \langle a_1 \bar{a}_2 a_3 \rangle = 0.33 \langle a_1 \bar{a}_2 \rangle & \langle \bar{a}_1 \bar{a}_2 a_3 \rangle = 0.44 \langle \bar{a}_1 \bar{a}_2 \rangle \\ \langle a_3 a_4 \rangle = 0.25 \langle a_3 \rangle & \langle \bar{a}_3 a_4 \rangle = 0.85 \langle \bar{a}_3 \rangle \end{array}$$

We also add the complementary consumers defined by Equation (10).

$$\begin{array}{ll} \langle \bar{a}_1 \rangle = 0.6 \langle T \rangle & \\ \langle a_1 \bar{a}_2 \rangle = 0.8 \langle a_1 \rangle & \langle \bar{a}_1 \bar{a}_2 \rangle = 0.7 \langle \bar{a}_1 \rangle \\ \langle a_1 a_2 \bar{a}_3 \rangle = 0.89 \langle a_1 a_2 \rangle & \langle \bar{a}_1 a_2 \bar{a}_3 \rangle = 0.78 \langle \bar{a}_1 a_2 \rangle \\ \langle a_1 \bar{a}_2 \bar{a}_3 \rangle = 0.67 \langle a_1 \bar{a}_2 \rangle & \langle \bar{a}_1 \bar{a}_2 \bar{a}_3 \rangle = 0.56 \langle \bar{a}_1 \bar{a}_2 \rangle \\ \langle a_3 \bar{a}_4 \rangle = 0.75 \langle a_3 \rangle & \langle \bar{a}_3 \bar{a}_4 \rangle = 0.15 \langle \bar{a}_3 \rangle \end{array}$$

Consider the first equation above, $\langle a_1 \rangle = 0.4 \langle T \rangle$. The CES consumer representing this relationship has an interest in the two goods $\langle a_1 \rangle$ and $\langle T \rangle$, with CES $\alpha$ coefficients of 0.4 and 1, respectively. In our computational market we endow the consumer with an amount $e = 10$ of each good, and set the global substitution parameter $\sigma$ to 50. The remaining consumers are instantiated in the same way.

For each node in the Bayesian network we add the arbitrage producers defined in Equation 11. In this example, we need to add producers only for node $a_4$, since the set $W_i$ is empty for $i \neq 4$.

$$\langle a_3 \rangle = \langle a_1 a_2 a_3 \rangle + \langle \bar{a}_1 a_2 a_3 \rangle + \langle a_1 \bar{a}_2 a_3 \rangle + \langle \bar{a}_1 \bar{a}_2 a_3 \rangle$$

$$\langle \bar{a}_3 \rangle = \langle a_1 a_2 \bar{a}_3 \rangle + \langle \bar{a}_1 a_2 \bar{a}_3 \rangle + \langle a_1 \bar{a}_2 \bar{a}_3 \rangle + \langle \bar{a}_1 \bar{a}_2 \bar{a}_3 \rangle$$

This collection of goods, consumers, and producers forms a complete MarketBayes economy. We have implemented this example in our market-oriented programming environment, WALRAS, which provides some general facilities for specifying computational markets (Wellman 1993). Given the specification of agents and goods, the WALRAS distributed bidding protocol attempts to find a competitive equilibrium via an asynchronous, iterative, price-adjustment process (Cheng and Wellman 1996). In the WALRAS bidding protocol, each agent submits demand functions for each of the goods they are interested in to the auctions for the respective goods. The auction then sets the price so as to clear its market. When the prices change, the agents may submit new bids, and the process iterates. For this example economy, the MarketBayes prices indeed converge correctly to within 0.001 of the correct probabilities.

We can also use the market results to recover probabilities of propositions that are not explicitly represented as goods in the system. Since the conditional probabilities in a Bayesian network capture the complete joint distribution, the probability of any propositional expression can be recovered through additions and multiplications of MarketBayes prices. For example, $\langle \bar{a}_1 a_3 \rangle$ can be computed by summing $\langle \bar{a}_1 a_2 a_3 \rangle + \langle \bar{a}_1 \bar{a}_2 a_3 \rangle$. For simple summations like this, we can generate an explicit arbitrageur to produce the desired good. However, determining the probabilities of general expressions and conditionals may require new types of agents, or even off-line calculations using the prices of existing goods.

## 7  DISCUSSION

In this paper, we have presented a specific technique for mapping a Bayesian network to a market system where prices are probabilities. Our market model encodes uncertain propositions as goods, conditional probabilities as consumers, and the laws of probability as arbitrage producers. We have shown that the competitive equilibrium of our system is unique, and corresponds to the joint probability distribution represented by the original Bayesian network.

We view the main contribution of this work to be an existence argument. We are fairly certain that there are other plausible mappings, some perhaps with advantages over the approach presented here (though we do believe that MarketBayes has some interesting features!).

The existence of a market model for Bayesian inference is the first step toward addressing some of the motivating research questions posed at the outset. We conjecture that market price systems will support a useful class of belief aggregation mechanisms, due to their high degree of decentralization, intuitive interpretation of interactions, and well-developed analytical theory. Within the MarketBayes framework, we can specify agents with differing beliefs (preference coefficient, $k$), rigidity of belief (substitution parameter, $\sigma$),[6] and relative importance (endowment, $e$). MarketBayes aggregates these consumers to derive an intermediate price; in ongoing work we are investigating the properties of this aggregation.

Market systems also have potential applications in distributed reasoning under uncertainty, where agents cannot be centrally coordinated due to high communication costs, actual physical separation, security concerns, or other reasons. In a precise sense, the competitive mechanism uses minimal communication (prices) to achieve—in particular, well-characterized circumstances—Pareto optimal outcomes.

Finally, the MarketBayes model also demonstrates that independence structure in uncertain belief states can be exploited to reduce the number of markets necessary to constitute a complete configuration. This may prove advantageous for problems of resource allocation under uncertainty, where the general theory would require unthinkable numbers of securities.

### Acknowledgments

We are grateful to the anonymous reviewers and students in the UM Decision Machines Group for careful reading and helpful suggestions. This work was supported in part by Grant F49620-94-1-0027 from the Air Force Office of Scientific Research, and an NSF National Young Investigator award.

---

[6] Under a suitable interpretation of the contingent goods, $\sigma$ can be considered a risk aversion parameter, with risk neutrality at $\sigma = \infty$. The more risk averse the agent, the more it prefers to hedge its beliefs by refraining from bets with positive expected payoff.